\documentclass[letterpaper]{aa}
\usepackage{graphicx}
\usepackage{rotating,subfigure,amssymb}
\usepackage{natbib}
\usepackage{subfigure}
\usepackage{txfonts}
\def\xmm{{\it XMM-Newton\/}}
\def\etal{et al.\ }
\def\betamod{$\beta$-model}

\def\kT {{\rm k}T}

\def\rs {r_{\rm s}}
\def \rv {r_{200}}

\begin{document}
     \title{Further evidence for a merger in Abell 2218 from
       an {\it XMM-Newton} observation}

     \author{G.W. Pratt, H. B\"ohringer and A. Finoguenov
     }
     \offprints{G.W. Pratt, \email{gwp@mpe.mpg.de}}

     \institute{MPE Garching, Giessenbachstra{\ss}e., 85748 Garching, Germany}
     \date{Received 4 June 2004; accepted 22 December 2004}

\abstract{The galaxy cluster Abell 2218, at $z=0.171$, is well-known
for the discrepancy between mass estimates derived from X-ray and
strong lensing analyses. With the present \xmm\ observation, we are
able to trace the gas density and temperature
profiles out to a radius of $\sim 1400~h_{70}^{-1}$ kpc (approximately
the virial radius of the cluster). The overall surface brightness 
profile is well fitted over three orders of magnitude with a simple
\betamod\, with a core radius of $0\farcm95$ and $\beta = 0.63$. The
projected temperature profile declines steeply with radius (by $\sim
50\%$), and is well described by a polytrope with parameters $t_0 =
8.09$ keV and $\gamma=1.15$. The temperature map
shows a pronounced peak in the central arcminute, where the
temperature rises by a factor of two (from $\sim 5$ to $\sim 10$
keV). The mass profile, calculated assuming hydrostatic equilibrium
and spherical symmetry, is best fitted with a King approximation
to an isothermal sphere, implying a dark matter distribution with a
central core, in contrast with the cusped cores
found in more obviously relaxed clusters. The X-ray
mass is approximately two times less than the strong lensing mass at
$r \sim 80~h_{50}^{-1}$ kpc, although the agreement between X-ray and weak
lensing mass measurements at larger radius ($r \sim 400~h_{50}^{-1}$
kpc) is slightly better. While the X-ray total mass estimates can vary
by 30 per cent depending on the mass model, all measurements are
significantly lower than the corresponding total mass from optical
measurements. Given the X-ray results indicating considerable
disturbance of the intracluster gas, leading to a probable violation
of the assumption of hydrostatic equilibrium, and the observed
substructure in the optical, suggesting a line-of-sight merger, it is
unlikely that the different mass estimates of this cluster can be
reconciled, at least with standard modelling assumptions. 

 }

     \authorrunning{G.W. Pratt et al.}
      \titlerunning{An \xmm\ observation of A2218} \maketitle
%

\begin{figure*}
\begin{centering}
\includegraphics[scale=1.,angle=0,keepaspectratio,width=\columnwidth]{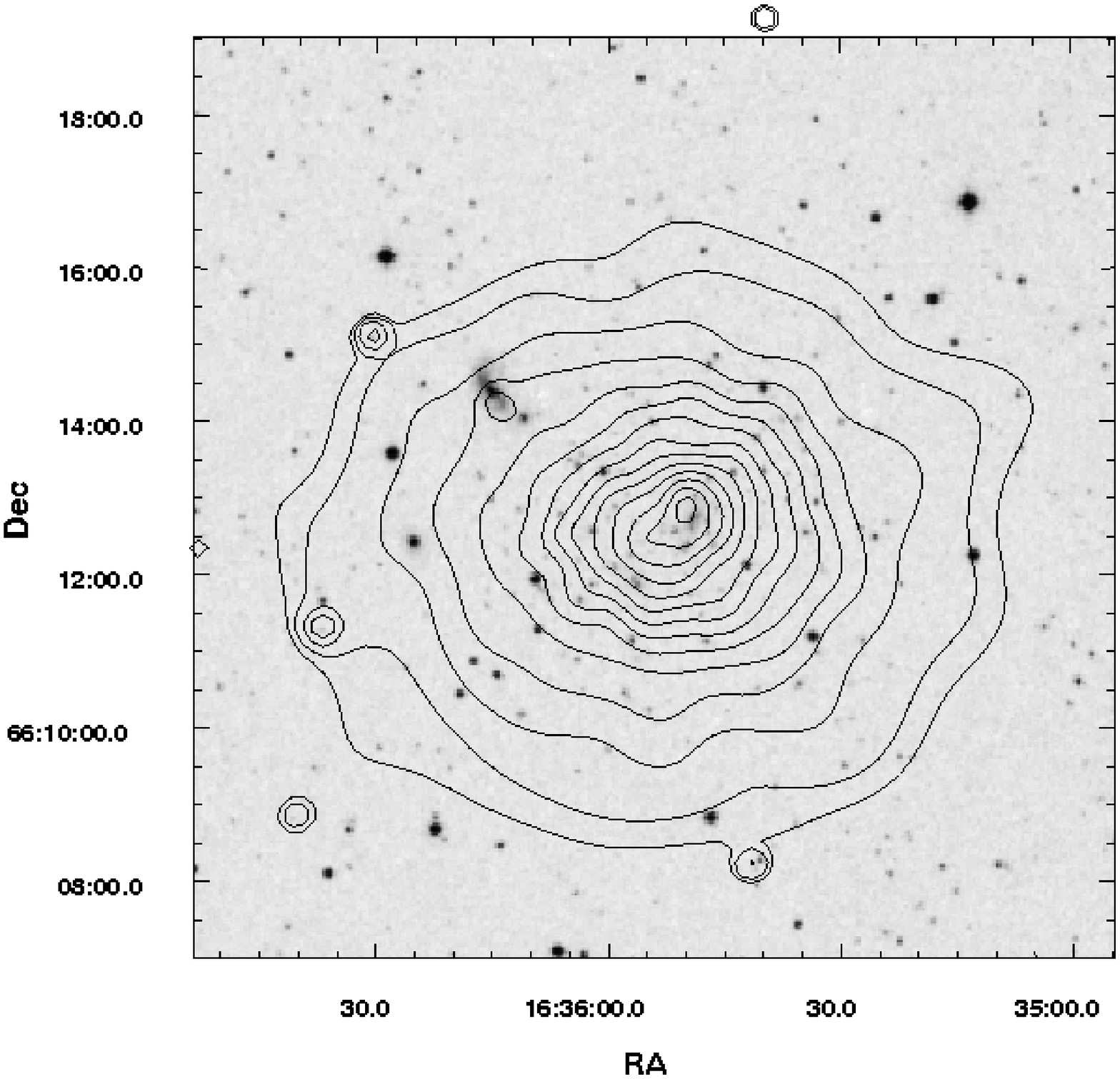}
\includegraphics[scale=1.,angle=0,keepaspectratio,width=\columnwidth]{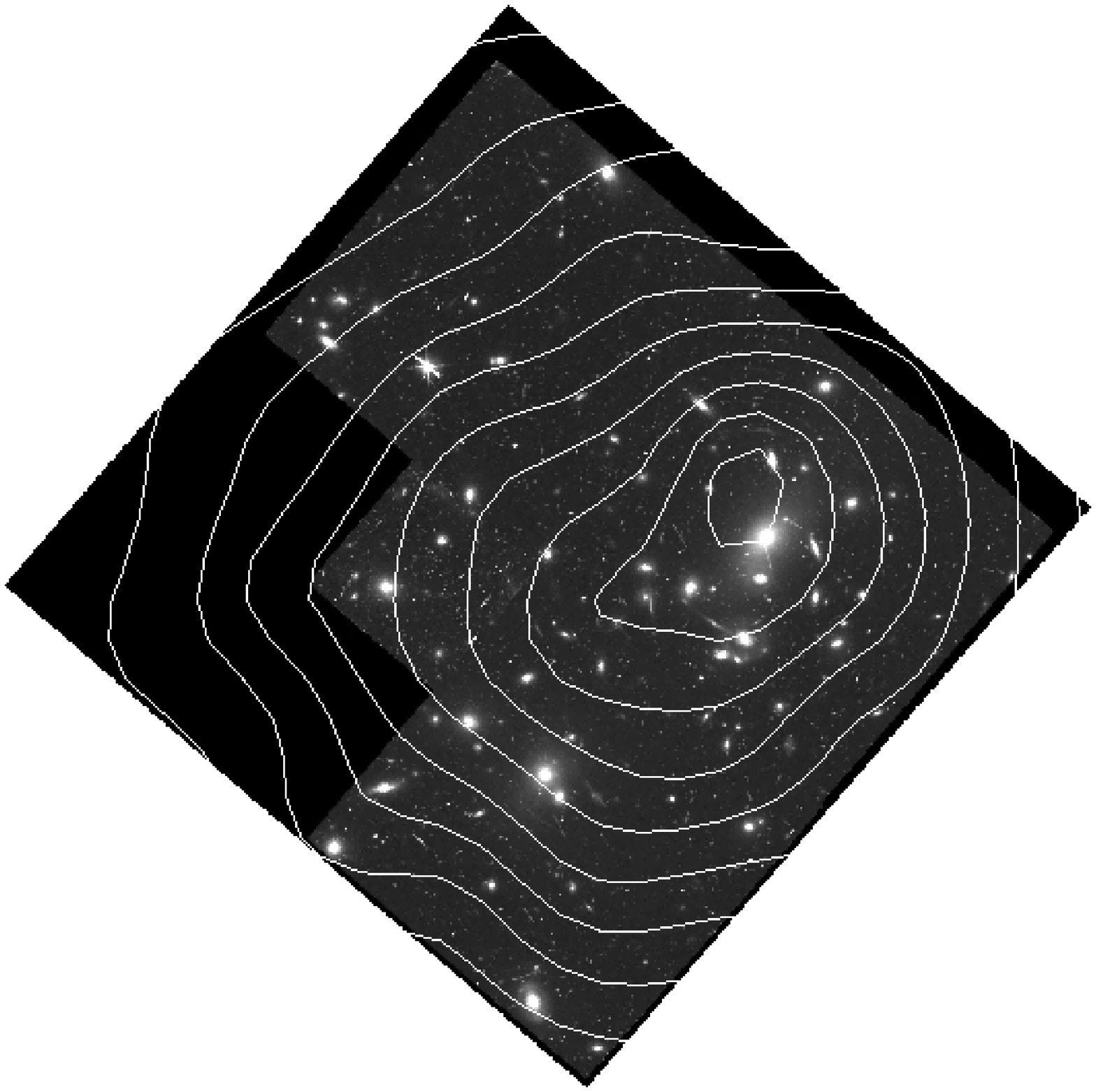}
\caption{{\footnotesize The left panel shows a large-scale XMM/DSS
    overlay image. The (log-spaced) contours are from the
    adaptively-smoothed, non-background subtracted, EMOS1+EMOS2
    [0.5-2.0] keV image. The right panel shows an XMM/HST overlay
    image with the same contours.  }}\label{fig:optX} 
\end{centering}
\end{figure*}

\section{Introduction}

In what has effectively become the standard scenario for the formation
of structure in the Universe, galaxy clusters are born in the collapse
of initial density fluctuations and grow under the influence of
gravity. They are the nodes of cosmic structure, accumulating mass
through a combination of continuous accretion of smaller subunits
along the filaments between clusters, interrupted by the occasional
violent merger. Cosmological simulations show that, in a given
cosmological volume, the sample of clusters will span the range from
those which are relatively relaxed to those which are clearly in the
midst of a major merger \citep[e.g.,][]{asc}, not surprising given the
hierarchical scenario.  

Given a sufficiently large sample of galaxy clusters, constraints can
be put on cosmological parameters. For instance, constraints on
$\sigma_8$ and $\Omega$ can be found by observing the cluster mass
function and its evolution (\citealt{perren}; see \citealt{henry2} for
a recent review). Cluster masses can be estimated using traditional
baryonic mass tracers such as the velocity dispersion of the galaxies
\citep[e.g.,][]{getal98}, or the X-ray emission of the intracluster
medium \citep[e.g.,][]{rb}, provided some assumptions are made about
the relationship between the distribution of dark and luminous
matter. For example, the fact that a given cluster is roughly
spherical and in approximate hydrostatic equilibrium is a fundamental
assumption when the total mass is derived from the X-ray data. On the
other hand, gravitational lensing offers another view of the
distribution of matter, allowing the calculation of the projected mass
within a cluster, either from strong lensing at small scales or weak
lensing on larger scales. However, because lensing measures the
projected mass, strong lensing in particular is very sensitive to
additional matter along the line of sight \citep*{wamb}, and to the
detailed dynamical state of the cluster \citep{torri}. Indeed, it has
been suggested that strong lensing arcs occur predominantly in
clusters which are dynamically more active than typical clusters
\citep{bs}.  

Cluster mergers release vast amounts of energy into the X-ray emitting
gas of the intracluster medium (ICM), leading to significant
departures from hydrostatic equilibrium. Strong variations are
produced in observable characteristics such as density, and, since
much of the energy is dissipated in shocks, temperature and
entropy. Temperature variations following a merger can survive $\sim
4-6$ times longer than density substructure, and so the temperature
distribution is a strong diagnostic of the dynamical state of a given
cluster \citep[e.g.,][]{mark3}. 

A2218 was one of the first targets for a detailed mass study comparing
X-ray and weak lensing data \citep{squires}, which concluded that
these mass estimates were in agreement (see also
\citealt{allen}). However, the strong lensing mass (estimated using
the remarkable arcs at a distance of $\sim 80
~h_{50}^{-1}$ kpc) was not in agreement with the X-ray estimate
\citep[e.g.,][]{mb}. {\it ROSAT\/} studies suggested complicated X-ray
structure near the core \citep{mark,nb99}; recent {\it Chandra\/}
observations have revealed this structure in greater detail
\citep{mach}. This has given rise to the general conclusion that the
cluster is dynamically active, which is also supported by the presence
of mass substructure in subsequent lensing studies (the latest being
\citealt{smith}), and indeed substructure in the galaxy distribution
in the optical \citep{gm}. The recent {\it Chandra\/} temperature map
of the central regions shows quite strong temperature variations
\citep{gov}. The cluster also hosts a weak radio halo \citep{gf}.  

With the present \xmm\ observations we derive gas density and
temperature profiles out to large radius, generate a large scale
temperature map, and, assuming hydrostatic equilibrium and spherical
symmetry, calculate a mass profile of the cluster. As we will see, the
assumptions inherent in the mass profile calculation are probably
violated, at least in the cluster centre.

The cluster lies at a redshift of $z=0.171$. In our chosen cosmology
($H_0 = 70$ km s$^{-1}$ Mpc$^{-1}$, $\Omega_m = 0.3$ and
$\Omega_{\Lambda} = 0.7$), 1\arcmin\ corresponds to $\sim 175$
kpc. Errors are quoted at the $1 \sigma$ level unless otherwise stated.

\section{Observations and data reduction}

A2218 was observed for $\sim 18$ ks on 2002 September 28. The
observation showed no soft proton flares in the high
energy ($> 10$ keV) light curves. There is a slight ($\sim 15$ per
cent) increase in the broad-band [0.3 -10.0 keV] count rate at the
beginning of the observation, lasting $\sim 2$ ks. Since the increase
is very weak and of short duration, and has no effect on the derived
normalisation at high energy, we have used the entire observation.

Except where noted, in the following analysis we use EMOS PATTERNs 1-12
and EPN PATTERN 0 events. To correct for vignetting, the photon
weighting method as implemented in the SAS task {\tt evigweight} was
applied to each event file. The blank-sky event lists of \citet{rp}
were used as background files, with these event 
lists being subjected to the same screening criteria, PATTERN
selection and vignetting correction as the observation event
files. 

\section{Gas density distribution}

\subsection{Morphology}

An XMM/DSS overlay image is shown in Fig.~\ref{fig:optX}. The cluster
exhibits slightly elliptical, symmetric X-ray emission at large
scale. Near the core, the X-ray 
emission becomes elongated in the SE-NW direction, which is the same
orientation as the line joining the two brightest cluster galaxies. As
can be seen in the XMM/HST overlay image, the peak of the [0.5-2.0]
keV band X-ray brightness is slightly offset from the cD galaxy. Both
the cD and the [0.5-2.0] keV X-ray peak are offset from the
large-scale centre of symmetry. Since the complicated X-ray morphology
of the central regions has been extensively examined using both {\it
  ROSAT\/} HRI \citep{mark} and {\it Chandra\/} \cite{mach}, we will
not discuss this further.  

\subsection{Surface brightness profile}
\label{sec:sbprof}

After identification and exclusion of 28 serendipitous sources, an
azimuthally averaged surface brightness profile was generated from
source and background event files of each camera in the 0.3-3.0 keV
band. Events were extracted in circular annuli centred on the emission
peak of the cluster. Background subtraction for each camera was
undertaken as described in \citet{pa02}. The EMOS and
EPN profiles are in excellent accord and thus the individual camera
profiles were coadded, and the total profile was then binned to a
$S/N$ ratio of $3\sigma$ above background.  This profile is shown in
Figure~\ref{fig:sbprof}. Cluster emission is 
detected out to slightly more than $11\arcmin$, corresponding to
slightly less than 2 Mpc for our chosen cosmology. For comparison, the
ROSAT profile of \citet{cann} was detected out to $\sim 12\arcmin$.

\subsection{Density profile modelling}
\label{sec:gasden}

\begin{figure}
\begin{centering}
\includegraphics[scale=1.,angle=0,keepaspectratio,width=\columnwidth]{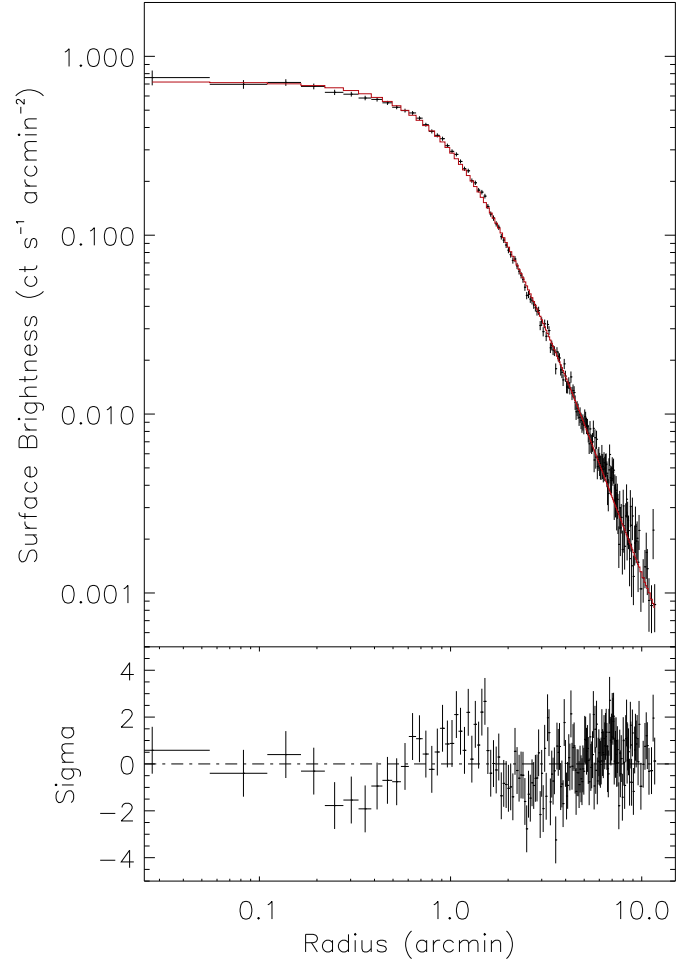}
\caption{{\footnotesize The combined EPIC 0.3-3.0 keV surface
    brightness profile of A2218. The profile is background subtracted
    and corrected for vignetting. The solid line is the best fitting
    \betamod. See Table~\ref{tab:sectors} for model
    details.}}\label{fig:sbprof} 
\end{centering}

\end{figure}

In order to calculate the gas density profile, we fitted the surface
brightness profile with analytic models. The fitting process uses
simplex minimisation and takes into account projection effects in the
line of sight. The models we use are convolved with the \xmm\ PSF and
binned into the same bins as the observed profile.  A standard
isothermal \betamod\ is an acceptable fit to these data ($\chi^2 =
208.8$ for 168 degrees of freedom), and the values found for $\beta$
and $r_c$ are in good agreement with ROSAT-derived values (e.g.,
\citealt{bh, squires, mark, cann}). In Figure~\ref{fig:sbprof} the
best-fitting 
model is shown compared to the surface brightness profile. In
Table~\ref{tab:sectors} we give the \betamod\ parameters for this fit
as well as for surface brightness profiles extracted in $90^{\circ}$
sectors defined anticlockwise from the positive Y-axis. In all
instances bar the $90^\circ - 180^\circ$ sector, a \betamod\ is an
adequate fit to the surface brightness. As a further check we
calculated surface brightness profiles centred on the broad-band X-ray
peak found in the {\it Chandra} analysis of \citet{mach}. These
profiles yield best-fitting \betamod\ parameters
which are in agreement with those in Table~\ref{tab:sectors}. 

On close examination of Fig.~\ref{fig:sbprof}, it can be seen that the
\betamod\ is an excellent description of the data over three orders of
magnitude in surface brightness. Related to this is the lack of
evidence for a turnover in the surface brightness at very large radius
(c.f., \citealt{vikh}), although this latter point is
highly sensitive to the exact background subtraction method. 

\begin{table}
\begin{minipage}{\columnwidth}
\caption{{\footnotesize \betamod\ fits for various sectors of A2218,
    errors are 90 per cent confidence. Sectors are defined
    anticlockwise from the positive Y-axis.}}\label{tab:sectors} 
\centering
\begin{tabular}{l l l l}
\hline
\multicolumn{1}{l}{Sector} & \multicolumn{1}{l}{$\beta$} & \multicolumn{1}{l}{$r_c$} & \multicolumn{1}{l}{$\chi^2$/dof}\\

\multicolumn{1}{l}{ } & \multicolumn{1}{l}{} & \multicolumn{1}{l}{ (arcmin) } \\
\hline

Full    & $0.63^{+0.01}_{-0.01}$ & $0\farcm95^{+0\farcm02}_{-0\farcm02}$ & 208.8/168 \\
\hline
0-90    & $0.63^{+0.01}_{-0.01}$ & $1\farcm04^{+0\farcm05}_{-0\farcm05}$ & 140.0/127 \\
90-180  & $0.61^{+0.01}_{-0.01}$ & $0\farcm98^{+0\farcm04}_{-0\farcm04}$ & 275.6/148 \\
180-270 & $0.74^{+0.03}_{-0.02}$ & $1\farcm21^{+0\farcm07}_{-0\farcm06}$ & 86.9/96 \\
270-360 & $0.66^{+0.01}_{-0.01}$ & $0\farcm87^{+0\farcm04}_{-0\farcm04}$ & 161.8/115 \\

\hline

\end{tabular}
\end{minipage}
\end{table}

\section{Spatially resolved spectroscopy}

In the following, the background spectra were normalised using the
count rates in the source-free [10.-12.] keV (EMOS) and [12.-14.] keV
(EPN) bands. The normalisation factors used were 1.09, 1.11 and 1.15 for
EMOS1, EMOS2 and EPN, respectively.

\subsection{Global spectrum}
\label{sec:ktglob}

We extracted a global spectrum from all events within a circle of
radius $5\farcm1$ to allow for direct comparison with the {\it
  Chandra\/} results. We fitted the data using a simple single
temperature absorbed MEKAL model with the redshift fixed at the
average redshift of the cluster ($z=0.17$). 

It was found that naive fitting of the spectra in the band [0.3-10.0]
keV with the absorption fixed at the galactic level ($3.24 \times
10^{20}$ cm$^{-2}$) produced results where the EPN temperature was
significantly lower than that found by the EMOS. Conversely, if the
absorption was let free, then the EPN absorption was significantly
lower than that found by the EMOS (which was in good agreement with the
galactic value). We thus adopted the {\it ad hoc\/} strategy discussed
in \citet{pa02}, fixing the low-energy cutoff at 0.6
keV for the EMOS and 1.0 keV for the EPN, which leads to good agreement
between EPIC temperatures with the absorption fixed at the galactic
value. 

The global temperature in this case is $6.63 \pm 0.27$, with an
abundance of $0.13 \pm 0.04$ solar (90\% confidence), the latter
measured relative to \citet{agr}. This is in good
agreement with previously determined values, such as the {\it Ginga\/}
analysis of \citet{mch}, the ASCA analyses of
\citet{ml},\citet{allen} and \citet{cann}, and the {\it Chandra\/}
  analysis of \citet{mach}.  

\subsection{Radial temperature profile}
\label{sec:tprof}

We then calculated a simple projected radial temperature
profile. After exclusion of point sources, spectra were extracted in
10 annuli centred on the X-ray emission peak and extending to the
radius of maximum detection of the surface brightness profile (i.e.,
0\arcmin - 11\arcmin).  The spectra were binned to $3\sigma$ to allow
the use of Gaussian statistics. EPIC spectra were then
simultaneously fitted in the [1.-10.] keV band with a single
temperature MEKAL model with free abundances and the absorption fixed
at the galactic value. We choose a relatively high low-energy cutoff
to avoid problems related to the variation of the soft X-ray
background with position on the sky.
 
While we have already normalised the background using the count rates
in the source-free high energy bands, background subtraction is
critical in spectroscopic studies at the 
lowest surface brightness levels, and can have significant effects on
the radial profile at high radii. We thus re-fitted the spectra with
the background at $\pm 10$ per cent of nominal, the variation seen in
the typical quiescent background level. The temperatures of all
annuli agree within the $1 \sigma$ error bars whatever the
normalisation.  

\begin{figure}
\begin{centering}
\includegraphics[scale=1.,angle=0,keepaspectratio,width=\columnwidth]{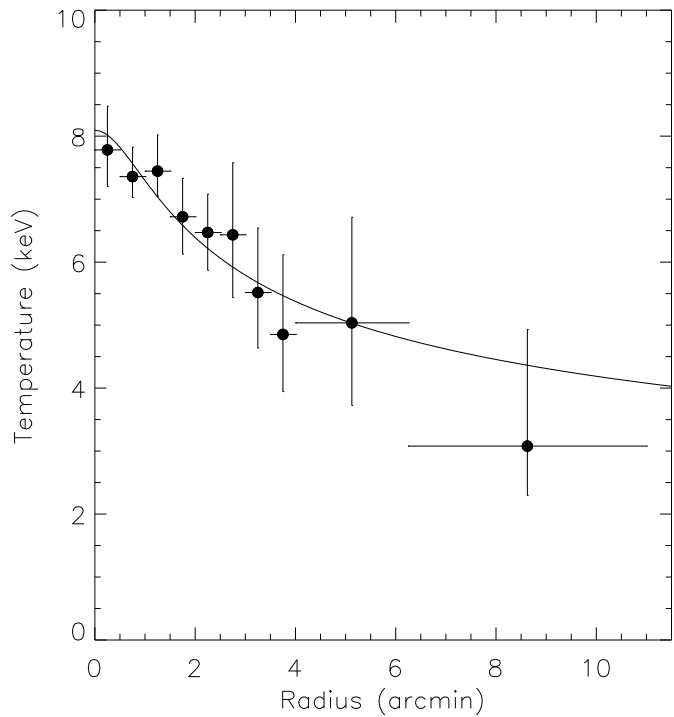}
\caption{{\footnotesize The projected radial temperature profile of
    A2218. Error bars are upper (lower) limits from fits with
    background normalisation at +10 (-10) per cent of
    nominal. The solid line is the best fitting
    polytropic model to the data. See text for
    details. }}\label{fig:tprof}  
\end{centering}

\end{figure}

The projected temperature profile is shown in
Fig.~\ref{fig:tprof}. The error bars are the upper and lower limit
temperatures for each annulus taking into account variations of $\pm
10$ percent in the background normalisation. These uncertainties
are the total systematic plus statistical errors and should be viewed
as very conservative estimates. As expected, the 
external annuli exhibit the largest uncertainties; however, the
robustly determined lower limit of the temperature in the final
annulus is 2.3 keV. The temperature profile is in excellent agreement
with that found in the {\it ASCA} analysis of
\citet{cann} and the 
more recent {\it Chandra} analysis of \citet{ma2218}, but
extends to larger radius (albeit with large uncertainties).
 
A polytropic model, $T \propto \rho_{\rm gas}^{(\gamma
  -1)}$, with the gas density described with the \betamod\ discussed
above (Sect.~\ref{sec:gasden}), is a good description of the projected
temperature profile. The best-fitting values for the central
temperature and polytropic index are $t_0 = 8.09 \pm 0.23$ keV and
$\gamma = 1.15 \pm 0.02$, respectively; this model
is also shown in Fig.~\ref{fig:tprof}.

As a consequence of the high temperature, abundance measurements of
this cluster are poorly-constrained, but are consistent with being
flat at the average value ($\sim 0.2$) out to the detection limit.

\subsection{Temperature map}
\label{sec:tmap}

\begin{figure}
\begin{centering}
\includegraphics[scale=1.,angle=0,keepaspectratio,width=\columnwidth]{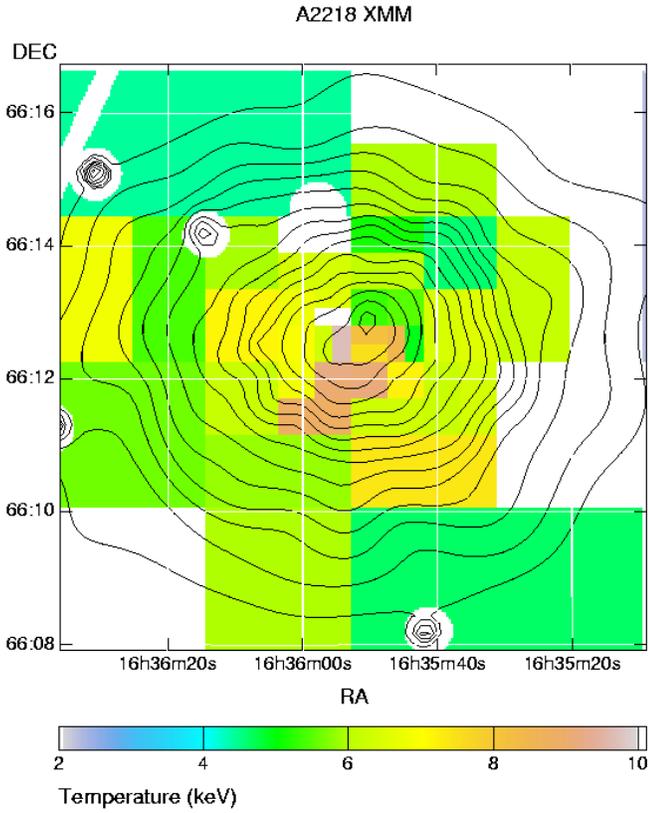}
\caption{{\footnotesize Adaptively-binned temperature map of
    A2218. The contours are the same as those in
    Fig.~\ref{fig:optX}.}}\label{fig:tmap}  
\end{centering}

\end{figure}

We next produced a temperature map using the adaptive binning
technique described in \citet{belsole1}. A full energy
band image with the sources excised was extracted from each camera,
and the camera images coadded. Starting from the largest spatial scale
(i.e., the entire image), the image was divided into cells. Each cell
is divided by four until a certain threshold criterion is met (for
this map, a total of 1000 counts in the full energy band prior to
background subtraction). Spectra are then extracted from the source
and background event lists with the resulting spatial binning, and
background subtracted. The same absorbed MEKAL model was fitted to
each spectrum in the bands described above (this time with the
abundance fixed to $0.2 Z_{\odot}$). The derived uncertainties are
dependent on the temperature. In building the temperature map shown in
Fig.~\ref{fig:tmap}, we only used cells with relative errors smaller
than 50 per cent. The
temperature structures in this map are very similar to those in the {\it
Chandra\/} map of \citet{gov}.

\section{Mass analysis}

\subsection{Deprojection of temperature profile and correction for PSF
  effects} 
\label{sec:deprojection}

In order to have the best estimate of the mass profile it is necessary
to take into account PSF and projection effects on the temperature
profile. Since correction for these effects using simultaneous
spectral fits under XSPEC tends to be unstable \citep{a478, pa04}, we
use the method described in \citet*{pap}. With the gas density
represented by the best fitting \betamod\ (Sect.~\ref{sec:gasden}), we
adopt a polytropic representation of the projected temperature profile
(Sect.~\ref{sec:tprof}), and convolve this with a redistribution
matrix which takes into account both PSF and projection effects. The
projected profile was randomly sampled 1000 times within the observed
errors, the gas density profile being allowed to vary within the
uncertainties of the best fitting \betamod. The final deprojected,
PSF-corrected temperature profile and associated 
errors were then the mean and standard deviation of the 1000 corrected
values at each point.

Fitting a polytropic model to the deprojected, PSF-corrected
temperature profile yields best-fitting values for the central
temperature and polytropic index of $t_0 = 8.66 \pm 0.1$ keV and
$\gamma = 1.16 \pm 0.01$, respectively.

\subsection{Total mass profile calculation}
\label{sec:mprof}

Combining the deprojected, PSF-corrected density and temperature
profiles  (Sects.~\ref{sec:gasden}
and~\ref{sec:deprojection}), we can derive a total mass profile under
the assumptions of hydrostatic equilibrium and spherical
symmetry. Given the observed temperature structure, this is probably
not a good assumption (certainly in the inner $1\farcm5$ or so), but
it is still 
interesting to see if X-ray and lensing mass estimates agree in the
external regions. The mass profile was calculated using the Monte
Carlo method described in \citet{pa03}; the resulting
total mass profile is shown in Fig.~\ref{fig:totmass}.  

For easier comparison with previous analyses, particularly results
from ASCA \citep{cann} and {\it Chandra} \citep{mach}, we
also show the mass profile obtained when the temperature profile is
assumed isothermal and the gas density is parameterised by a
\betamod. For this we assume a temperature of $\kT = 6.6$
keV (Sect.~\ref{sec:ktglob}) and the best fitting \betamod\ in
Sect.~\ref{sec:gasden}. The mass profile obtained using this method is
slightly different from the one we have derived using spatially
resolved temperature information, being slightly lower in the inner
regions, and higher in the outskirts.

\subsection{Mass profile modelling}
\label{sec:mmod}

\begin{figure}
\begin{centering}
\includegraphics[scale=1.,angle=0,keepaspectratio,width=\columnwidth]{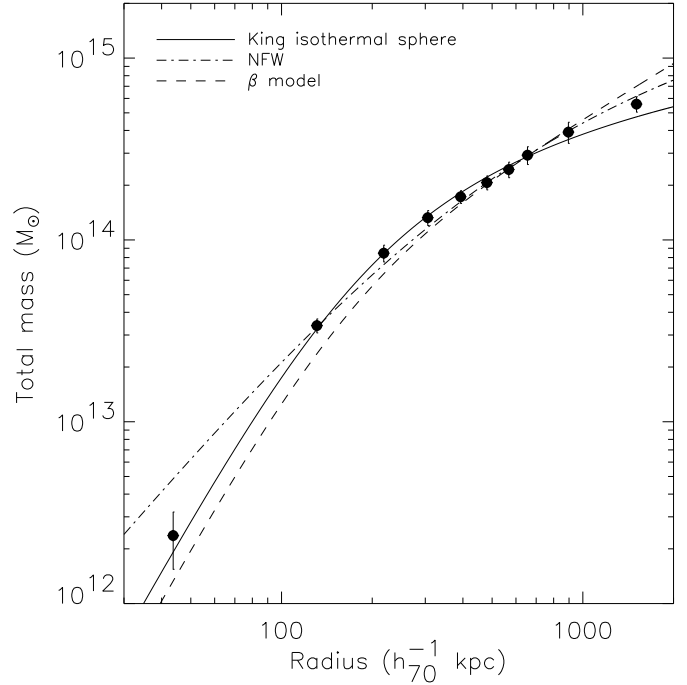}
\caption{{\footnotesize The total mass profile of A2218, calculated
    from the X-ray data assuming hydrostatic equilibrium and spherical
    symmetry. The dashed line is the isothermal \betamod\ mass,
    assuming a constant temperature of $\kT = 6.6$ keV
    (Sect.~\ref{sec:ktglob}).The solid line is the best fitting
    isothermal sphere model (King approximation) for the total mass
    profile; also shown (dot-dash line) is the  best fitting NFW mass
    model to the entire profile (see Table~\ref{tab:massprofs} for
      details).}}\label{fig:totmass}  
\end{centering}

\end{figure}

Since the mass profile is defined to very good precision, it is
interesting to compare it with theoretical expectations. We thus tried
fitting the mass profile with functional forms for the dark matter
density, these being the King approximation to an isothermal sphere
\citep[e.g.,][]{sarazin} and the NFW profile \citep*{nfw97}, given by: 

\begin{eqnarray}
M(r) & = & 4\pi \rho_{\rm c}(z) \delta_c \rs^3 m(r/\rs) \\
\delta_c & = & \frac{200}{3} \frac{c^3}{[ln (1+c) - c/(1+c)]} \\
m(x, {\rm NFW}) & = & \ln (1+ x) - x (1 + x)^{-1} \\
m(x, {\rm iso}) & = & \ln [x + (1 + x^2)^{1/2}] - x(1 + x^2)^{-1/2},
\end{eqnarray}

\noindent respectively, where we have followed \citet*{edm}, in
expressing the normalisation of both mass models through the
concentration parameter, $c$. The King profile is clearly preferred in
terms of $\chi^2$, although the removal of the central mass point
leads to an acceptable fit for the NFW mass model (see
Table~\ref{tab:massprofs} for details). The best-fitting mass profile
models are compared to the data in Fig.~\ref{fig:totmass}.    

\begin{table}
\begin{minipage}{\columnwidth}
\centering
\caption{ NFW and King isothermal sphere fits to the total mass
  profile of A2218. Errors are $1 \sigma$.} 
\begin{tabular}{ l l l l l l l}
\hline
\multicolumn{1}{l}{ Parameter }  & \multicolumn{2}{c}{ NFW Model } &
\multicolumn{1}{l}{ King model } \\
\multicolumn{1}{l}{ }  & \multicolumn{1}{l}{ Full } &
\multicolumn{1}{l}{ Central point } & \multicolumn{1}{l}{ } \\
\multicolumn{1}{l}{ }  & \multicolumn{1}{l}{ } &
\multicolumn{1}{l}{ removed } & \multicolumn{1}{l}{ } \\

\hline
$c$ & $4.96^{+0.43}_{-0.42}$ & $5.78^{+0.60}_{-0.54}$ &
$9.24^{+0.52}_{-0.50}$\\ 
$r_{\rm s}$ ($h_{70}^{-1}$ kpc)& $345^{+39}_{-34}$ & $289^{+38}_{-33}$
& $146^{+11}_{-10}$ \\
$r_{200}$ ($h_{70}^{-1}$ kpc)& 1715  & 1669 & 1525\\
$M_{200}~(M_{\odot})$ & $6.8 \times 10^{14}$ & $6.3 \times 10^{14}$
& $4.8 \times 10^{14}$\\
$\chi^2/{\nu}$ & 14.2/8 & 2.5/7 & 4.8/8\\
\hline
\end{tabular}
\label{tab:massprofs}
\end{minipage}
\end{table}



\section{Discussion}

\subsection{Temperature, entropy and pressure structure}

It has been known for some time that the central regions of
this cluster show signs of being morphologically disturbed in X-ray
\citep{mark,nb99,mach}. The spatially resolved temperature maps from this
work and from {\it Chandra} \citep{gov} provide further
evidence, if any were needed, that A2218 is undergoing (or has
undergone) a merger. The
centrally-peaked radial temperature profile has been resolved into a
small region of significantly hotter gas in the core. The temperature
varies by a factor of two in the central arcminute. The hottest
parts of the cluster are only approximately spatially coincident with
the weak radio halo \citep{gov}. Comparing the HST galaxy distribution
(Fig.~\ref{fig:optX}) 
and the temperature map (Fig.~\ref{fig:tmap}), the hottest region is
spatially coincident with the two main galaxies, and thus with the two
main mass peaks found in lensing analyses, but is offset from the
X-ray surface brightness maximum. 

We have also investigated the structure in entropy and pressure 
using masks created to resolve the observed fluctuations in (i) the
temperature and surface brightness, and (ii) the entropy and
pressure. Both masks have 16 regions, 8 regions covering the
central $1\farcm5$, and 8 covering larger scales of $1\arcmin
-5\arcmin$. The method of mask generation and the calculation of the
deprojected pressure and entropy in each region is described in
\citet*{henry}. The 
fluctuations on these scales are found to be statistically significant
and amount to $(11 - 13) \pm 2$ per cent in entropy and $14 \pm 3$ per
cent in pressure, where the two values given for the entropy show the
slight dependence on the mask. Note however that the results obtained
using the two different masks are consistent. The scatter in the
entropy and pressure in the central arcminute is $17\pm3$ per cent and
$17\pm5$ per cent, respectively.

The temperature structure of a cluster has often been used to learn
more about the merger kinematics in interacting clusters
\citep[e.g.,][]{msv,belsole1,belsole2}. This is very difficult in the present
case, since, at least in X-ray, the merger geometry is not
obvious. In particular, the X-ray surface
brightness does not show more than one peak and is fairly symmetric at
large radii.

\subsection{Entropy profile}

\begin{figure}
\begin{centering}
\includegraphics[scale=1.,angle=0,keepaspectratio,width=\columnwidth]{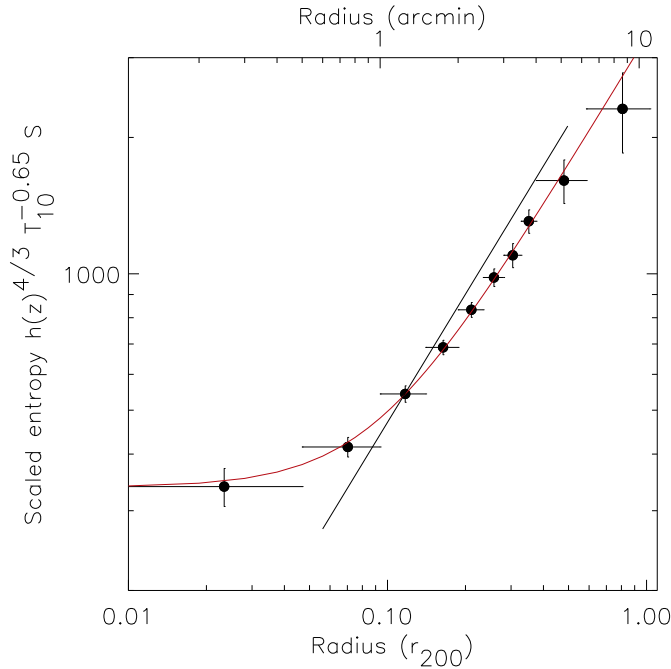}
\caption{{\footnotesize The entropy profile of A2218. The radius has
    been scaled to $r_{200}$ using the $r_{200}$--$T$ relation of
    \citet{app}. The entropy is scaled using the
    empirical scaling of \citet[][$S \propto h(z)^{-4/3}
    T^{0.65}$]{psf}. The line following the points is the
    entropy profile obtained from the best-fitting analytical models
    to the deprojected, PSF-corrected gas density and temperature
    profiles. The straight line is the mean scaled entropy profile found
    by \citet{pa04}.}}\label{fig:eprof} 
\end{centering}

\end{figure}

The entropy profile of the cluster shown in Figure~\ref{fig:eprof} was
calculated using the now-customary entropy definition, $S = \kT
n_e^{-2/3}$, where we have used the best-fitting \betamod\ for the gas
density (Sect.~\ref{sec:gasden}) and the deprojected, PSF-corrected
temperature profile (Sect.~\ref{sec:deprojection}). Uncertainties due to gas
density profile modelling are negligible compared to the temperature
measurement errors. The radius has been scaled using the observed 
$r_{200}$--$T$ relation of \citet*{app}\footnote{The relation is
  $\log{r_{200}} = (2.81 \pm 
0.02) + (0.57 \pm 0.03) \log{T}$, giving $r_{200}=1892~h_{70}^{-1}$
kpc for a global temperature of 6.6 keV (Sect.~\ref{sec:ktglob}).},
and the entropy has been scaled  
using the relation found by \citet[][]{psf}\footnote{$S \propto hz^{-4/3}
    T^{0.65}$}. We also show an
analytically-derived profile 
obtained using the best fitting analytic description of the
temperature and density data. 

It is clear that the entropy profile of A2218 is rather flat in the
inner regions ($r \lesssim 200~h_{70}^{-1}$ kpc, or $\theta \lesssim
1\arcmin$), where the central entropy levels out to a value of $S \sim
200~h_{70}^{-1/3}$ keV cm$^{2}$ at $20 h_{70}^{-1}$ kpc. This
flattening is driven by the large core in the gas density. Because of
this, A2218 was found to be a significant outlier in the sample of
\citet{lpc}. In fact, the temperature map shows variations of a factor
of two (from $\sim 5$ to $\sim 10$ keV) in the central 
arcminute. The complex
entropy distribution in the central regions is smoothed out by the use
of annular regions to produce the profile.

From an analysis of five clusters, \citet{pa04} found approximately 20
per cent dispersion in the scaled entropy profiles of their
sample. Their mean scaled profile, obtained in the radial range [0.05
- 0.5] $\rv$ (Eqn. 4 of \citealt{pa04}) is also shown in
Fig.~\ref{fig:eprof}. Excluding the inner three points, a power-law
fit to the entropy profile of A2218 yields a slope of $\alpha = 0.80
\pm 0.07$, slightly shallower than the mean scaled profile, although
consistent within the $1 \sigma$ uncertainties. 
 
\subsection{Mass estimates}

\subsubsection{Mass profile: comparison to theoretical expectations}

It is becoming increasingly clear that the mass profiles of relatively
relaxed, hot clusters are well described by the NFW profile (e.g.,
\citealt{david01,asf01,abg02,pa02,lewis,a478,pap}). Such clusters
generally exhibit strongly peaked surface brightness profiles, and
regular density and temperature structure. In cases such as
these, the standard assumptions in the X-ray mass analysis
(hydrostatic equilibrium and spherical symmetry) are more
than likely warranted. 

An NFW model is not a good description of the total mass profile of
A2218, particularly in the inner regions, as expected given the
observed density and temperature structure.  While exclusion of the
central mass point improves the NFW fit considerably, the best fitting
mass model to the entire profile remains the King model, a consequence
more of the large core in the gas density than the radial temperature
distribution.

\subsubsection{Comparison with lensing}

\begin{figure}
\begin{centering}
\includegraphics[scale=1.,angle=0,keepaspectratio,width=\columnwidth]{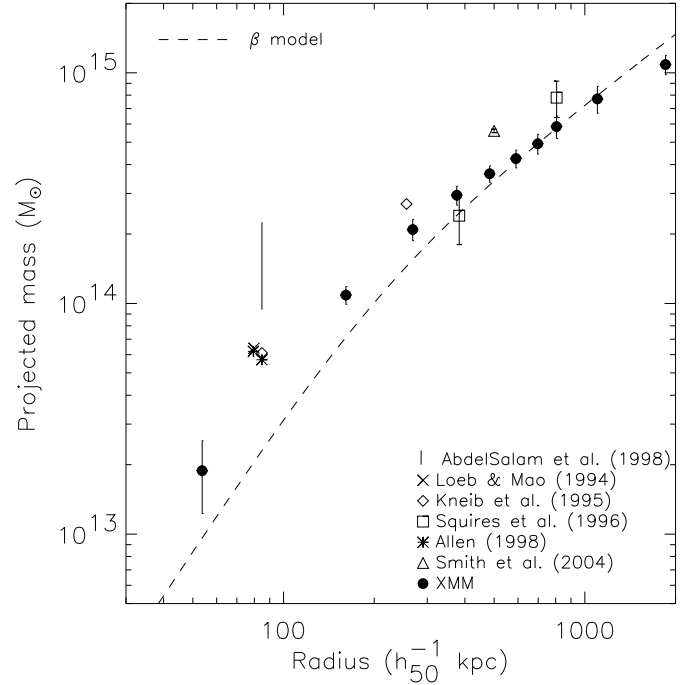}
\caption{{\footnotesize The projected mass profile of A2218. Filled
    circles with error bars are the XMM data points. The dashed line
    is the isothermal \betamod\ mass, assuming a constant temperature
    of 6.6 keV. Various other
    symbols indicate different lensing mass
    estimates.}}\label{fig:masscomp} 
\end{centering}

\end{figure}

The considerable disagreement between the masses derived from X-ray
and lensing analyses of this cluster has generated some debate in the
past \citep[e.g.,][]{mb,allen}. In Fig.~\ref{fig:masscomp} we show the
projected mass profile obtained from these \xmm\ data, converted to
the SCDM ($\Omega_0=1.0, \Omega_{\Lambda}=0.0, h=0.5$) cosmology used
in previous analyses. The profile is compared to the masses obtained
from strong lensing \citep{lm,kneib95,allen}, weak lensing
\citep{squires,smail} and combined lensing \citep{abdel,smith}.

Despite having a well constrained radial temperature distribution, the
X-ray mass is approximately two times less than the strong lensing
mass at $85~h_{50}^{-1}$ kpc.  At larger radius, the agreement between
mass estimates is better, although still not perfect. The observed
disagreement in the inner regions would be expected if either (i)
there was a smaller subcluster in the line of sight, or (ii) the
hydrostatic assumption is violated, possibilities we discuss in more
detail below.  

\subsubsection{Total mass estimate}

Depending on the adopted mass
model, the virial mass, $M_{200}$, varies by about 30 per cent (see
Table~\ref{tab:massprofs}).
\cite{app} have recently derived an X-ray $M$--$T$ relation derived from \xmm\
  observations of 10 clusters in the 2 -- 9 keV range. Since the
  methods detailed in the present paper are very
  similar to their analysis, it forms an ideal comparison sample.
The total mass, $M_{200}$, predicted by the $M$--$T$ relation of
\cite{app}, assuming a virial temperature of 6.6 keV
(Sect.~\ref{sec:ktglob}), is $8.5 \times
10^{14}~h_{70}^{-1}M_{\odot}$, about 25 per cent higher than the
highest mass estimate from mass profile modelling
(Table~\ref{tab:massprofs}).

We can also compare the mass estimates from the different X-ray mass
models with the optical virial mass estimates of \citet{gm}. We note
that the most distant galaxy from the cluster centre in Girardi \&
Mezzetti's analysis is only at $\sim 450 ~h_{70}^{-1}$ kpc, and so
there is a fair amount of extrapolation needed in their mass
estimate. These authors analyse 43 galaxies and find $M_{\rm v, opt} =
(2.1 \pm 0.5) \times 10^{15} ~h_{70}^{-1}~M_{\odot}$, for an
optically-derived virial radius of $2.3 ~h_{70}^{-1}$ Mpc. Our
best-fitting mass models give $8.6/ 4.7 \times
10^{14}~h_{70}^{-1}~M_{\odot}$ (NFW/King) for the same radius. This is
a very significant mass discrepancy. If there is indeed a line of sight
merger, this will artificially increase the optical galaxy velocity
dispersion if only one component is assumed, leading to a higher
optical mass estimate. 


\section{Summary and conclusions}

We have undertaken a detailed X-ray analysis of the well-known lensing
cluster A2218. In the inner 
regions ($\theta \lesssim 1\arcmin$, or $r \lesssim 200~h_{70}^{-1}$
kpc) the X-ray isophotes become elongated and oriented approimately
SW-NE, the same direction as the line joining the two brightest
cluster galaxies (which are also the mass peaks seen in lensing
studies). There is a marked offset between the position of the cD
galaxy and the X-ray brightness peak, as was seen in previous ROSAT
and {\it Chandra} studies \citep{mark,nb99,mach}. In agreement with
previous {\it Chandra} results \citep{ma2218}, we find that the
temperature profile decreases significantly from the centre. With the
present observation, we find that this decrease continues to the
virial radius of the cluster. From an adaptively-binned temperature
map, we find similar temperature structure to that found in the {\it
  Chandra} observation \citep{gov}. There is a pronounced peak in the
temperature in the inner $1\arcmin$, where the temperature rises from
$\kT \sim 5$ keV to $\kT \sim 10$ keV. The peak is not correlated with
any obvious surface brightness feature, and is offset from the surface
brightness maximum. The scatter in deprojected entropy and pressure in
the central arcminute is typically $\sim 17 \pm 5$ per cent.

The mass profile (determined assuming HE and spherical symmetry) is
best fitted with a King mass model, in contrast with the results from
more relaxed clusters. An NFW model is a particularly bad fit in the
central regions. The X-ray mass estimate does not agree with the mass
estimated from strong lensing arcs at $80~h_{50}^{-1}$, although the
agreement improves when comparing X-ray and weak lensing masses at
large radii. X-ray total mass estimates dependent on mass profile
fitting can vary by 30 per cent depending on the mass model
used. While the results must be viewed with caution, the total mass
measurement is 20-60 per cent smaller than the mass expected from the
$M$--$T$ relation. X-ray and lensing mass measurements are
significantly smaller than optical virial mass estimates by more than
a factor of two.

An important clue to the merger geometry in this system comes from
optical observations. The distribution of galaxy velocities in the
optical study of \cite{getal97} strongly indicates substructure in the
line of sight. Given this and the observed X-ray characteristics, it
is likely that A2218 is either an ongoing merger between a main
cluster and a smaller subcluster, or a single cluster in a more
advanced merger state. In either case, the central regions, at least,
of the cluster are unlikely to be in hydrostatic equilibrium. As a
result of this, X-ray mass estimates are highly
uncertain. Discrepancies between X-ray and lensing mass estimates due
to line of sight effects and/or mergers have also been discussed in
\citet[][A1698]{am} and \citet[][CL0024+17]{yyz}, and in the case of
A2218 itself, by \citet{mach}, in the context of the {\it Chandra}
observation.

Perhaps the most significant result from our analysis is the
calculation of a well-constrained radial temperature profile out
to approximately the virial radius of the cluster. It is unfortunate
in a way that A2218 is so obviously undergoing a merger event. In
other circumstances it would have been an ideal benchmark object for
mass measurements using the various methods in use today. 

\begin{acknowledgements}

GWP is grateful to J.-L. Sauvageot, M. Arnaud and E. Pointecouteau for
useful discussions. We thank the referee for comments which allowed us
to improve the paper. GWP acknowledges funding from a Marie Curie
Intra-European Fellowship under the FP6 programme (Contract
No. MEIF-CT-2003-500915). AF acknowledges support from BMBF/DLR under
grant 50 OR 0207 and partial support from NASA grant NNG04GF686. The
present work is based on observations obtained with {\it XMM-Newton},
an ESA science mission with instruments and contributions directly
funded by ESA Member States and the USA (NASA). This research has made
use of the NASA's Astrophysics Data System Abstract Service; the
SIMBAD database operated at CDS, Strasbourg, France; the High Energy
Astrophysics Science Archive Research Center Online Service, provided
by the NASA/Goddard Space Flight Center and the Digitized Sky Surveys
produced at the Space Telescope Science Institute.

\end{acknowledgements}

\end{document}